\documentclass[sn-mathphys-num]{sn-jnl}%

\usepackage{graphicx}%
\usepackage{multirow}%
\usepackage{amsmath,amssymb,amsfonts}%
\usepackage{amsthm}%
\usepackage{mathrsfs}%
\usepackage[title]{appendix}%
\usepackage{xcolor}%
\usepackage{textcomp}%
\usepackage{manyfoot}%
\usepackage{booktabs}%
\usepackage{algorithm}%
\usepackage{algorithmicx}%
\usepackage{algpseudocode}%
\usepackage{listings}%

\begin{document}

\title[Article Title]{Askaryan Radio Array: searching for the highest energy neutrinos}

\author*[1]{\fnm{Mohammad Ful Hossain} \sur{Seikh}}\email{fulhossain@ku.edu}


\affil*[1]{\orgdiv{Department of Physics \& Astronomy}, \orgname{University of Kansas}, \orgaddress{\city{Lawrence}, \postcode{66045}, \state{KS}, \country{USA}}}

\abstract{Searches for ultra-high energy ($E_\nu \geq 10$ PeV) cosmogenic and astrophysical neutrinos (UHENs) have been conducted by several experiments over the last two decades. The Askaryan Radio Array (ARA), located near the geographical South Pole, was one of the first two experiments that used radio antennas sensitive to orthogonal polarizations for detection of neutrino-induced Askaryan radiation. ARA comprises five independent autonomous stations, with an additional low threshold phased array merged with station 5, which were deployed at a depth of 100-200 m over the period 2012-2018, corresponding to a total livetime of more than 27 station years. In this article, we present a brief overview of the detector, its detection technique, and discuss a few of its major achievements with a focus on the current status of the array-wide UHEN search. We expect to produce the most sensitive results on the neutrino flux by any existing in-ice neutrino experiment below 1000 EeV energy.}


\maketitle

\section{Introduction}\label{sec1}
\begin{figure}[h]
		\centering
        \includegraphics[scale = 0.65]{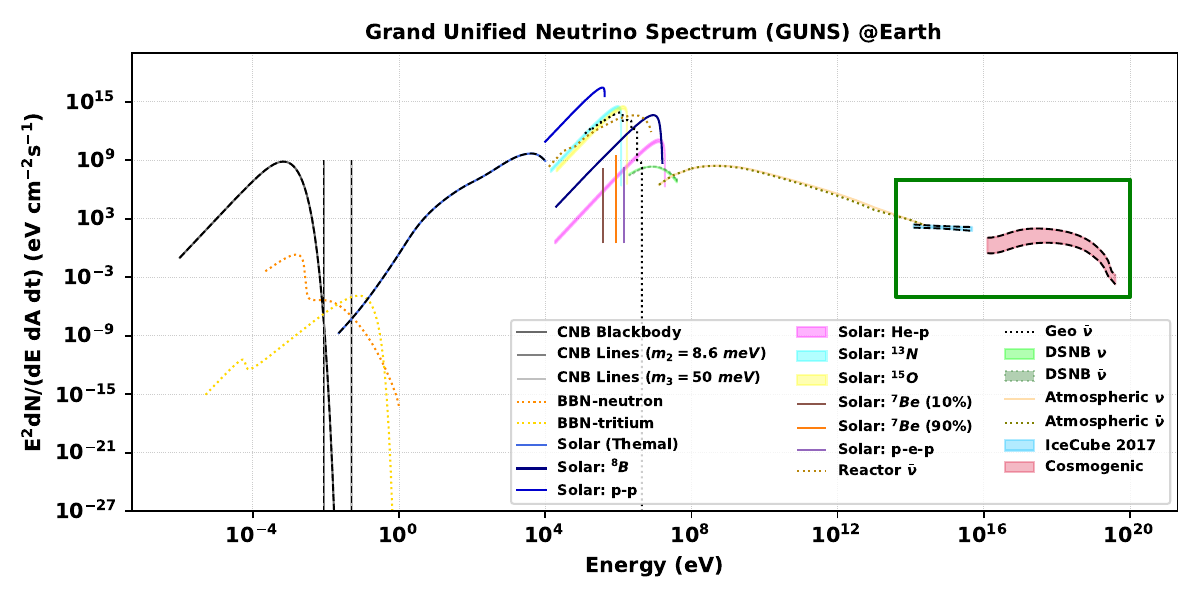}
		\caption{Neutrino spectrum at Earth integrated over directions and summed over flavors. Antineutrinos and  neutrinos are shown by dashed/dotted lines and solid lines, respectively. \textcolor{green!50!black}{Green} box shows the PeV and EeV neutrino flux. Adapted from \cite{3}.}
		\label{fig1}
\end{figure}

UHENs, carrying important source information over astronomical distance, are unique cosmic messengers. Their detection will open a new window to the high energy universe. Other UHE particles like Ultra-high cosmic rays (UHECR) are attenuated by interactions with the Cosmic Microwave Background (CMB) above the GZK limit (named after Greisen, Zatsepin and Kuzmin), $10^{19.5}$ eV \cite{1}\cite{2} and gamma rays above a few TeV annihilate with the CMB and Extragalactic Background Light. The advantage of almost zero annihilation while traversing cosmic distances, due to the weakly interacting nature of neutrinos, is somewhat offset by the minuscule flux of cosmic rays (refer to figure 30.1 in \href{https://pdg.lbl.gov/2024/reviews/rpp2024-rev-cosmic-rays.pdf}{PDG: Cosmic Rays}) and neutrinos  in the UHE regime as shown in figure \ref{fig1}). In addition, their low cross-section adds more challenges to their detection. To compensate the low flux and small cross section, detectors with large volumes are required \cite{15}. The $\mathcal{O}$(km) thick, naturally available ice sheets in Antarctica and Greenland are thus ideal places to deploy such detectors.
The primary Physics goal for the ARA detectors is to detect the Askaryan radiation induced by cosmogenic and astrophysical neutrinos in ice. 
\subsection{Askaryan Radiation}\label{subsec1.1}
\begin{figure}[h]
        \centering \includegraphics[scale = 0.2]{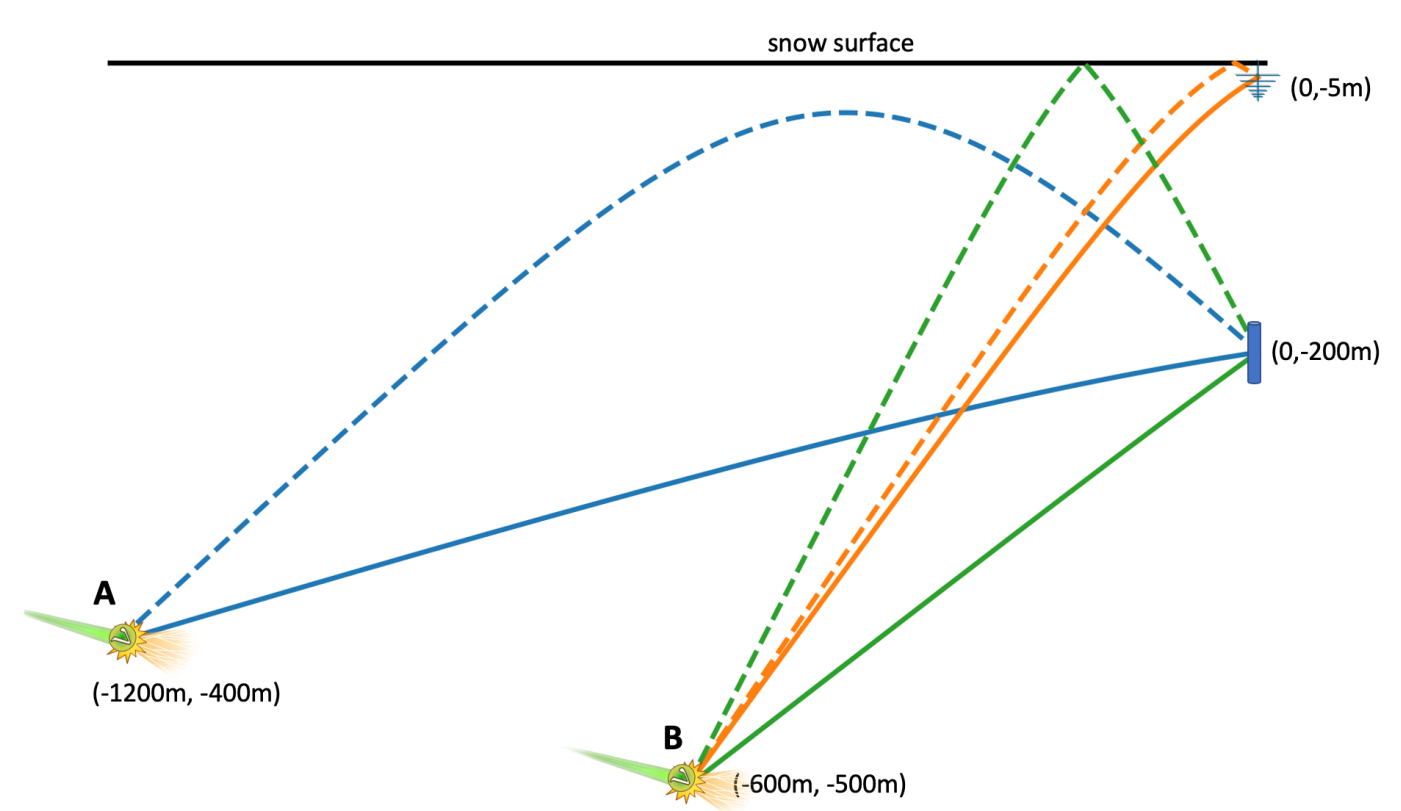}
		\caption{Bending of rays (direct, refracted, and reflected) due to the depth variation of the refractive index profile of South Polar ice. Source \cite{6}.}
		\label{fig2}
\end{figure}

In 1962, the Soviet-Armenian physicist Gurgen Askaryan postulated the emission of coherent radio waves from the interaction of cosmic rays in a dense media (now called `Askaryan radiation' \cite{4}). When an UHEN interacts with ice, it produces a cascade of secondary particles that develops a time varying negative charge-excess which gives rise to coherent radio emission. The charge excess is mainly caused by pair annihilation of positrons, and Bhabha, Moller, and Compton scatterings of electrons and positrons in the electromagnetic shower, as it interacts with the surrounding atomic electrons in ice. In deep ice with refractive index $n = 1.78$, the Cherenkov angle of radio emission is $\theta_c \simeq \cos^{-1}(1/\beta n) \sim 56^0$; incoming neutrino directions near the horizon are most favorable for the Cherenkov cone to be detected by in-ice antennas. Moreover, due to the exponential refractive index profile of ice \cite{5} with depth, the geometry of the ray bends down in the firn (See figure  \ref{fig2}) layer of ice, complicating reconstruction of the incoming neutrino direction. 
\subsection{Cosmogenic and Astrophysical Neutrinos}\label{subsec1.2}
The highest energy accelerators in our known universe are thought to be Active Galactic Nuclei (AGN) \cite{7} and Gamma Ray Bursts (GRBs) \cite{8} which are the primary sources of astrophysical neutrinos having PeV or higher energies. When an accelerated proton from these astrophysical objects interacts with the ambient matter (N), it produces pions which subsequently decay to neutrinos and gamma rays as shown in equations \eqref{eq1}, \eqref{eq2}, and \eqref{eq3}, 
\begin{align}
    p + N & \longrightarrow  \nonumber \pi + X~~~~(\pi = \pi^0, \pi^+, \pi^-)
     \\[1pt]
     \pi^0 & \longrightarrow \gamma + \gamma \label{eq1}\tag{1}\\[1pt]
     \pi^+ (\pi^-) & \longrightarrow \mu^+ (\mu^-) + \nu_\mu (\bar\nu_\mu) \label{eq2}\tag{2}\\[1pt]
    \mu^+ (\mu^-) & \longrightarrow e^+ (e^-) + \nu_e (\bar\nu_e) + \bar\nu_\mu (\nu_\mu)\label{eq3}\tag{3}.
\end{align}
Cosmogenic neutrinos with $E_\nu > 100$ PeV, on the other hand, are produced when UHECRs interact with CMB photons as shown in equation \eqref{eq4} 
\begin{align}
    p + \gamma_{CMB}  \xrightarrow[]{\hspace{2mm}\Delta^+\hspace{2mm}} 
    \begin{cases}
     p + \pi^0 \longrightarrow p + \gamma + \gamma\\
     n + \pi^+ \longrightarrow n + e^+ + \nu_\mu + \nu_e + \bar\nu_\mu    \end{cases} . 
\label{eq4}\tag{4}    
\end{align}
Detecting UHENs and UHECRs will not only help us understand their sources but also will contribute significantly towards multi-messenger astronomy.

\section{ARA Detector}\label{sec2}
\begin{figure}[h]
        \centering \includegraphics[scale = 0.25]{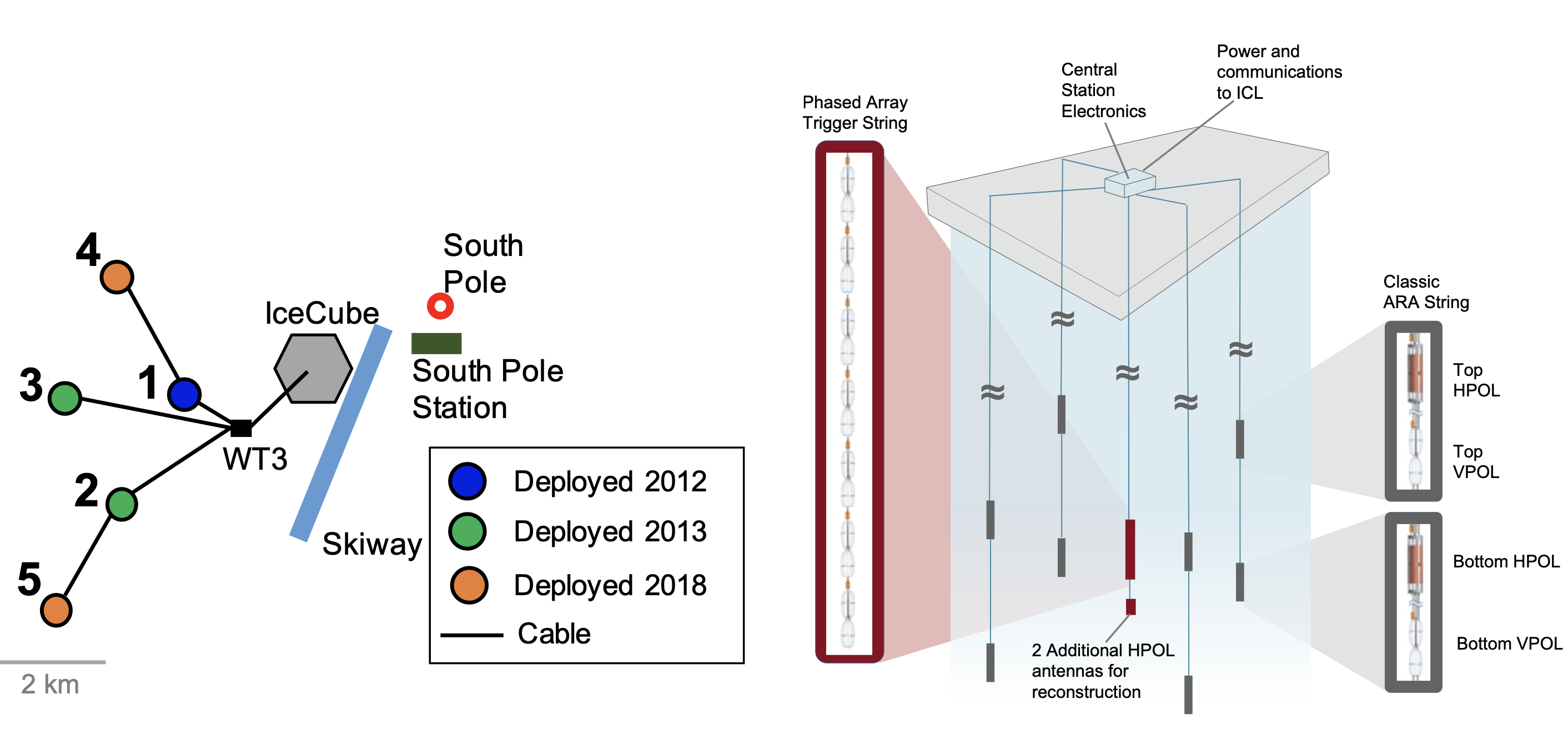}
		\caption{Schematic layout of ARA stations (left) and ARA station 5 (A5) + phased array antennas (right). ARA stations 1-4 (A1, A2, A3, A4) do not have the central phased array (PA) trigger string.}
		\label{fig3}
\end{figure}
ARA is an array of five autonomous stations (A1-A5) located about 2 km from the IceCube Neutrino Observatory. Before deploying the first ARA station, A1, a prototype TestBed was installed in January 2011 to test the viability and feasibility of the future planned ARA-37. It demonstrated the capability of a radio neutrino detector to constantly scan the in-ice ambient thermal noise and, if deployed on a large scale, detect cosmogenic neutrinos \cite{9}. Shortly afterwards, during the summer season of 2011-2012, A1 was deployed, although it started consistently taking data only in early 2014. The initial plan was to deploy the receiving antennas (Rx) beneath the firn ice at depths of at least 100 m; due to drilling difficulties, the in-ice antennas of this station could only be installed at depths $<100$ m. In  subsequent years, A2 and A3 were deployed at a depth from 170 to 190 m. A4 and A5 were deployed after 2018. All five ARA stations (ARA5) have similar geometry and instrumentation. Each station has a trapezoidal grid of four Rx strings with 16 radio antennas (distributed across 4 strings, each with 2 Vertically Polarized (VPol) and 2 Horizontally Polarized (HPol) antennas). In addition, 2 strings of transmitting (Tx) antennas, known as calibration pulsers (CP) or simply `calpulsers', each with 1
VPol and 1 HPol antennas are placed near the station center and at a right angle configuration from the central station electronics, as shown in figure \ref{fig3}. In A1-A3, we also deployed four surface antennas (SA) to detect cosmic rays but a hardware issue that restricted the bandwidth resulted in dropping the SAs in the design of later stations.

\subsection{Antennas}\label{subsec2.1}
\begin{figure}[h]
        \centering \includegraphics[scale = 0.4]{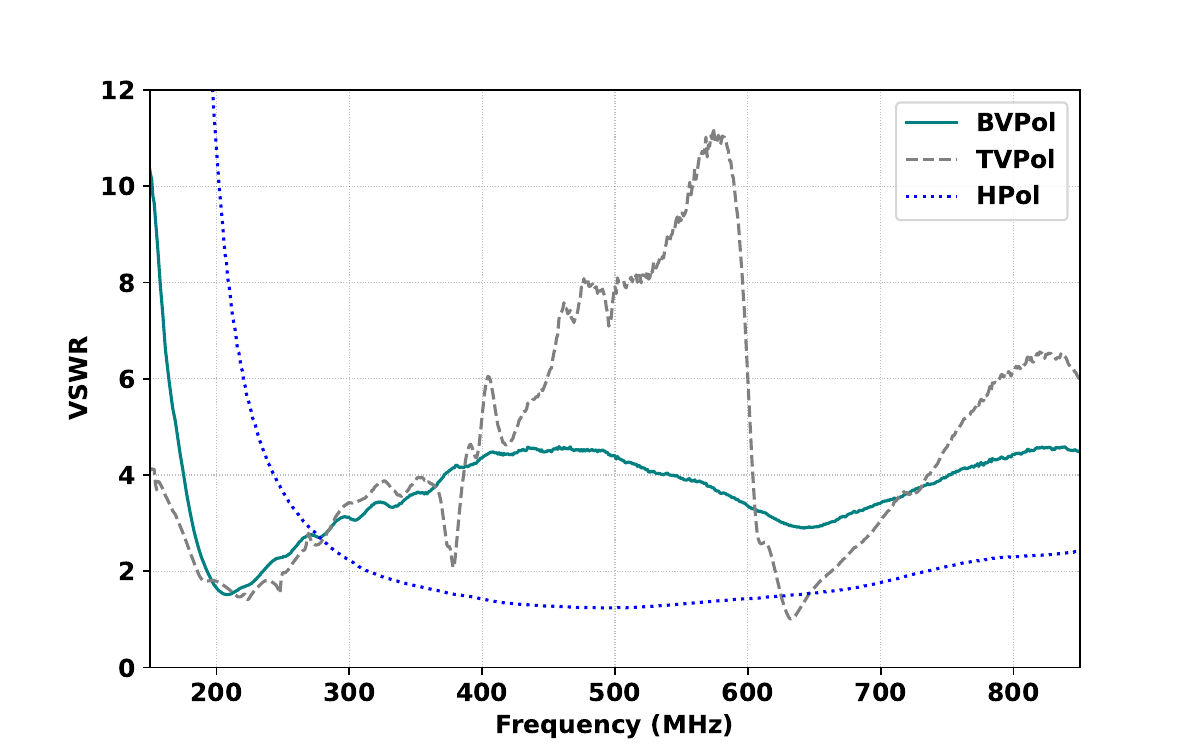}~\includegraphics[scale = 0.044]{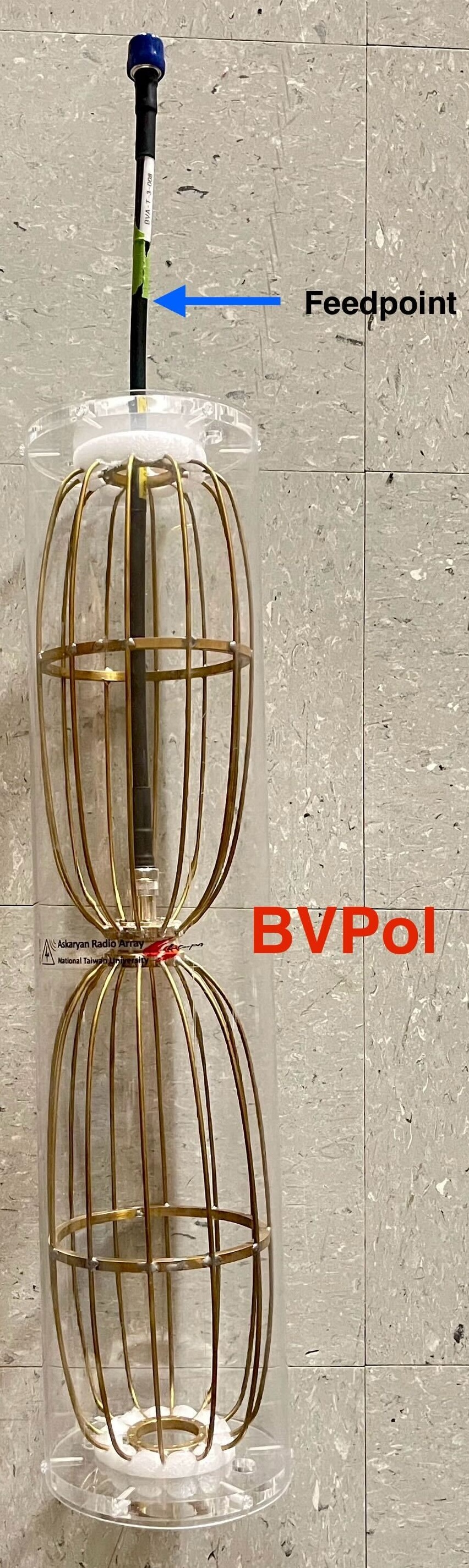}~\includegraphics[scale = 0.042]{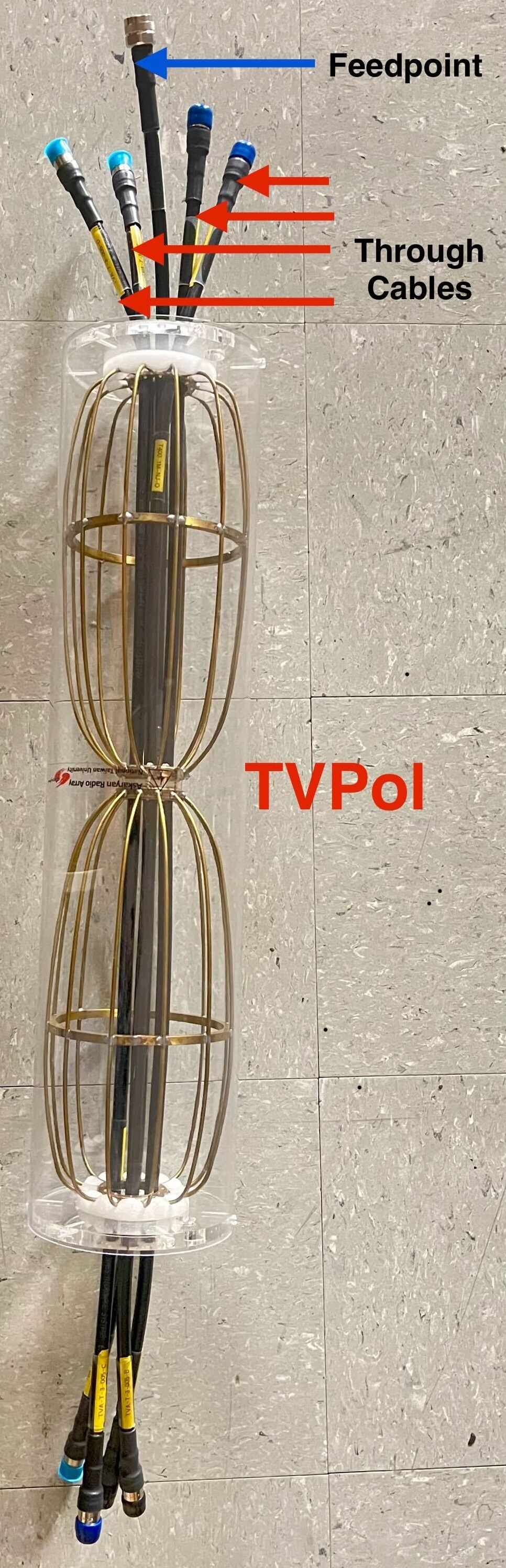}~\includegraphics[scale = 0.04]{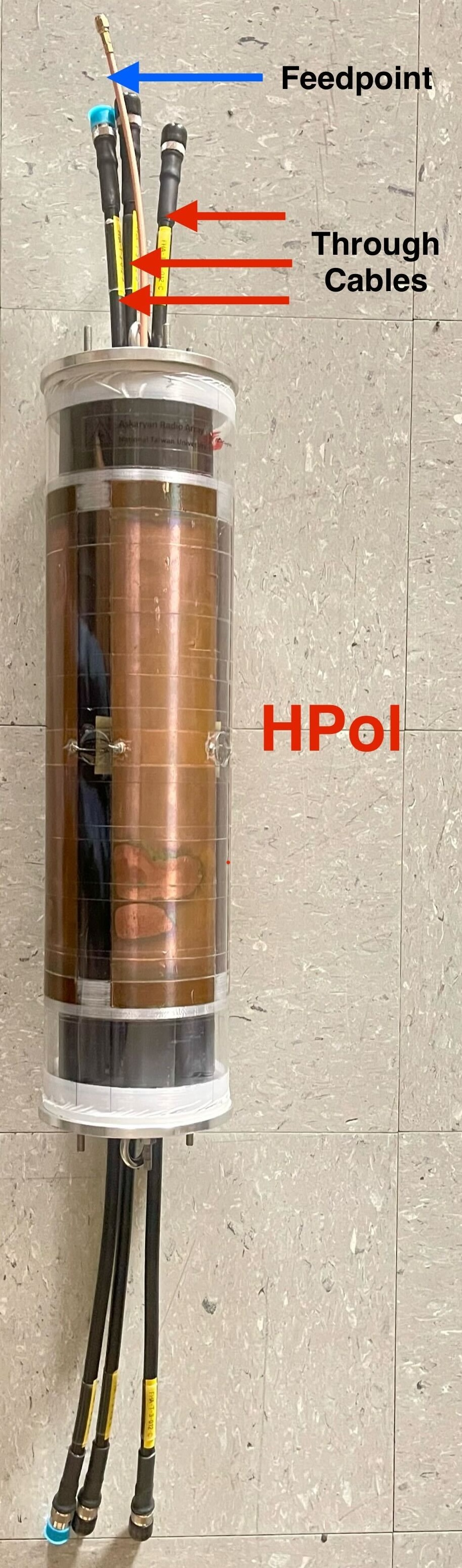}
		\caption{Voltage Standing Wave Ratio (VSWR) of the ARA antennas (left). Notice that HPol has 3, TVPol has 4 and BVPol has no through cables.}
		\label{fig4}
\end{figure}

ARA uses both VPol and HPol antennas to maximize the chance of detecting UHENs and UHECRs. In a string, from top to bottom, there is a top pair HPol (THPol) and VPol (TVPol), separated by 2 m. About 15-20 m below the top pair, another pair of bottom HPol (BHPol) and VPol (BVPol) are installed; see station string geometry in figure \ref{fig3}. Therefore, a number of through cables are passed through most of the antennas. The ARA antenna through-cables are therefore configured in two different ways: the ferrite-loaded quadslot HPol (either top and bottom with three through cables) and the birdcage VPols (top with four through cables and bottom with none). We found that these through cables make a noticeable difference in the antenna characteristics. For instance, the power reflection due to the impedance mismatch of the antenna and connecting transmission lines, measured via the Voltage Standing Wave Ratio (VSWR) (figure \ref{fig4}), is significantly worse in the TVPols due to the four extra through cables, despite the fact that the TVPol and BVPol are otherwise identical antennas. We note that the resonance frequency and first harmonics coincide in the top and bottom VPols due to their identical lengths and construction. 

\subsection{Signal Chain}\label{subsec2.2}
Signals received by the in-ice antennas reach the surface Data Acquisition (DAQ) via the `signal chain' (SC) hardware. The ARA SC is designed to provide a net power gain of about 75 dB as can be seen in figure \ref{fig9}. Some part of the signals intercepted by the antennas is lost due to impedance mismatch and, if off boresight, diminished by the zenith dependent antenna directivity (see figure \ref{fig5}). To ensure that the signal is large enough to register above the noise floor of the digitizer (typically 1 mV), a low noise amplifier (LNA with about 40 dB gain) is connected immediately after the antenna and a filter-module. The filter module, consisting of a band-pass and band-stop (notch) filter, is used to exclude unwanted out-of-band backgrounds, as well as narrow-band surface communications at 450 MHz respectively. After propagating  through a highly shielded LMR-series transmission line for a few meters, the electrical signal is converted to an optical signal and then conveyed to the surface DAQ through a Radio Frequency over Fiber (RFoF) module. Another transducer converts the optical signal back to an electrical signal, after which second-stage amplifiers provide another 40-50 dB gain. Before entering the DAQ box, a band-pass filter restricts the signal band to the interval 150-850 MHz. The signal chain thus boosts the received signal (as well as thermal noise) from $\mu$V scale to mV range.  

\begin{figure}[H]
        \centering \includegraphics[scale = 0.44]{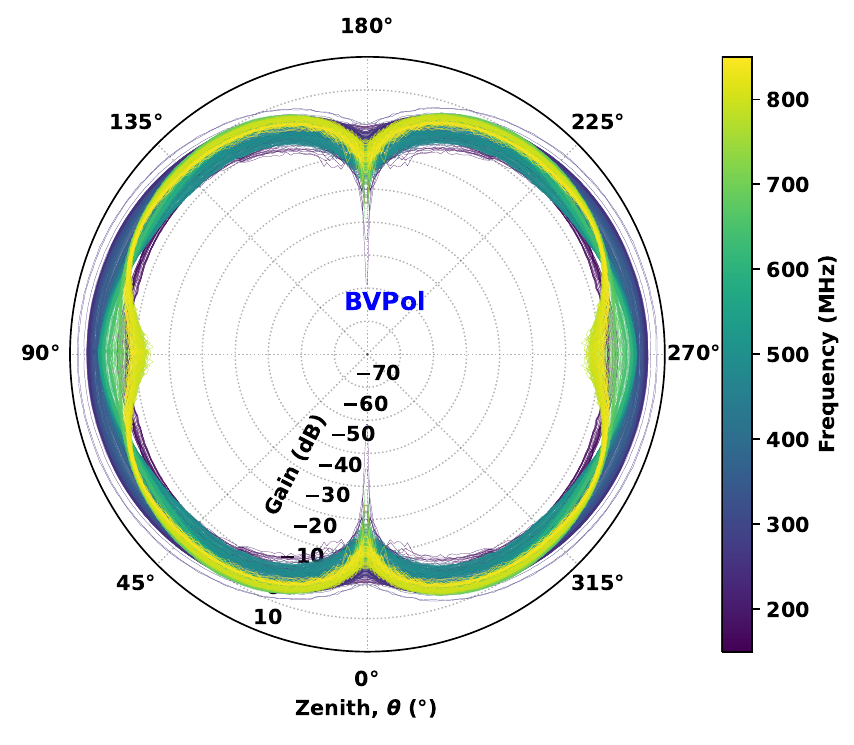}~\includegraphics[scale = 0.44]{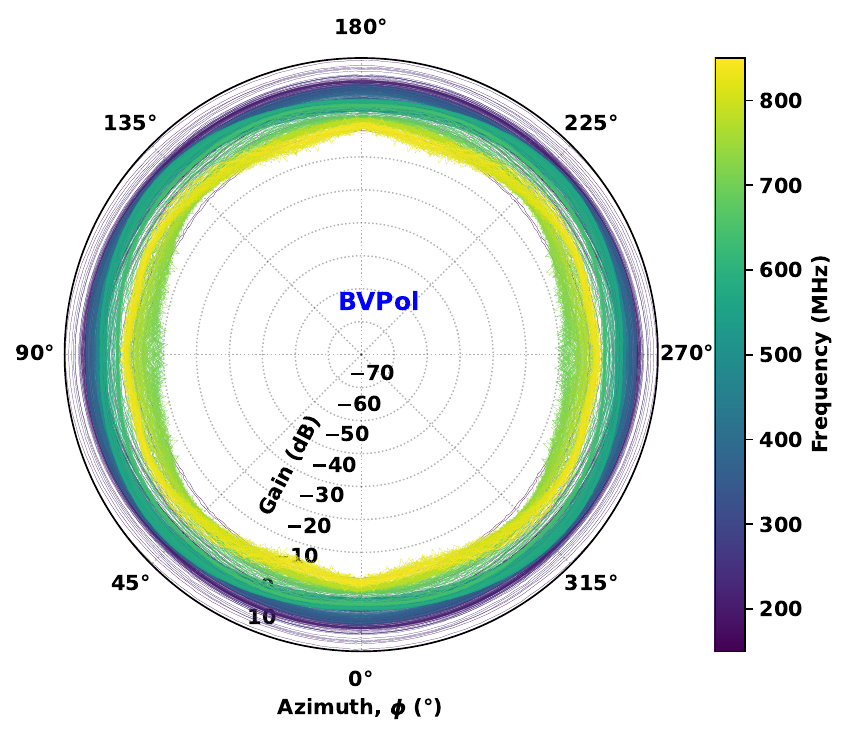}\\
		\includegraphics[scale = 0.44]{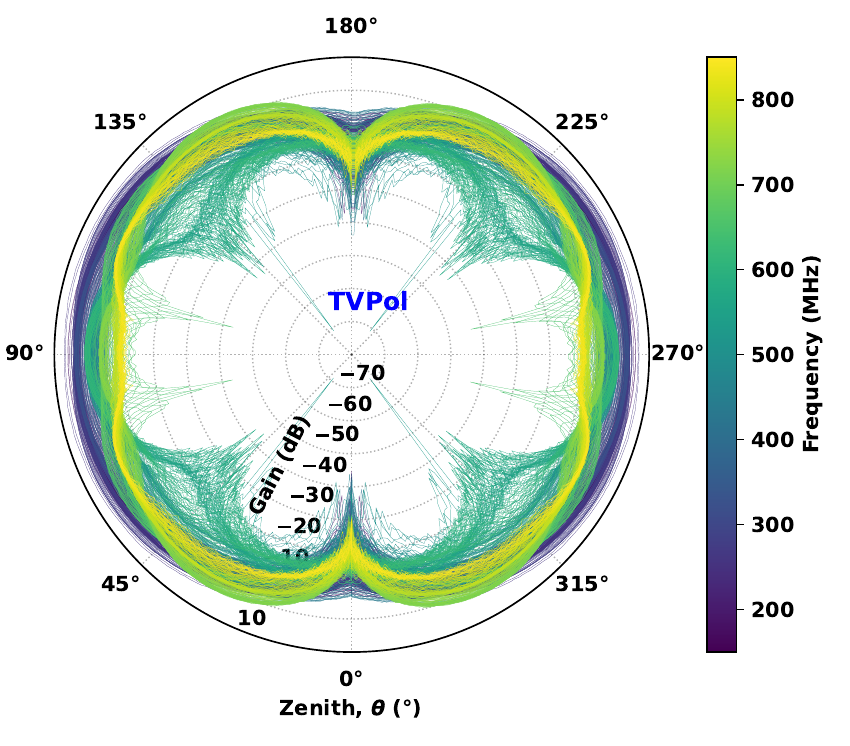}~\includegraphics[scale = 0.44]{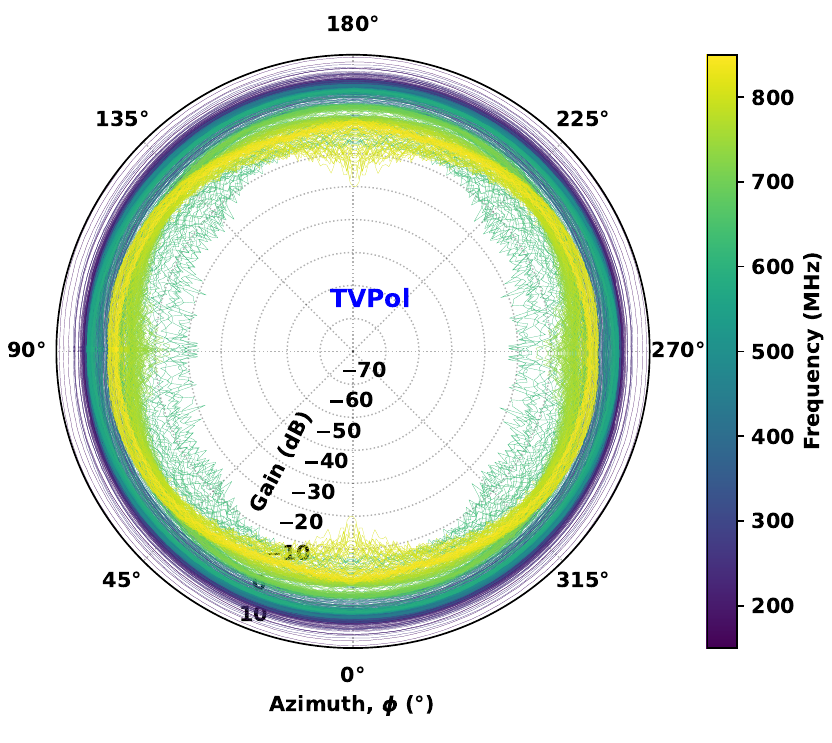}\\
        \includegraphics[scale = 0.44]{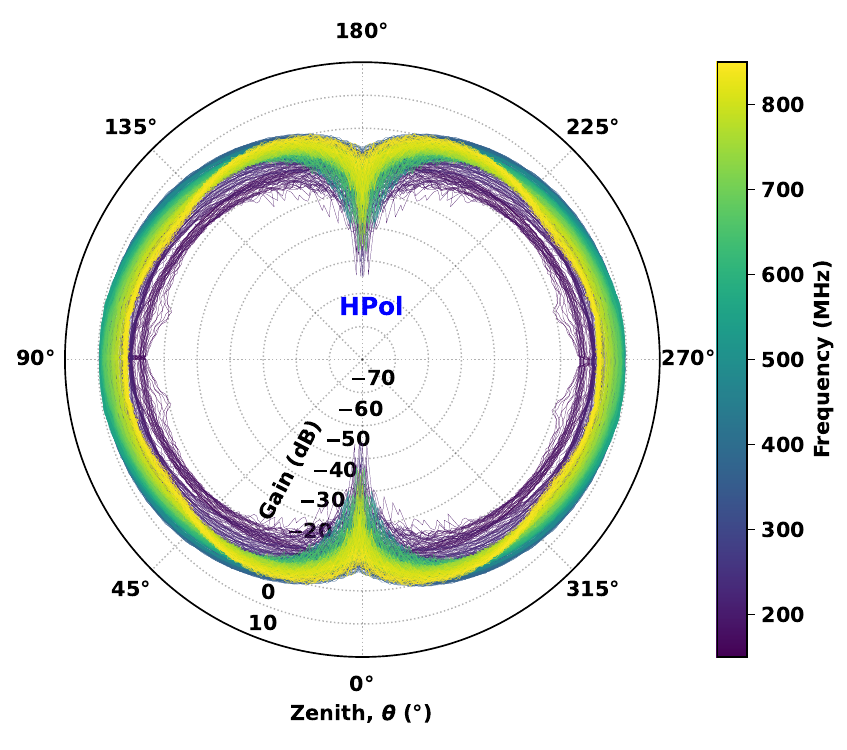}~\includegraphics[scale = 0.44]{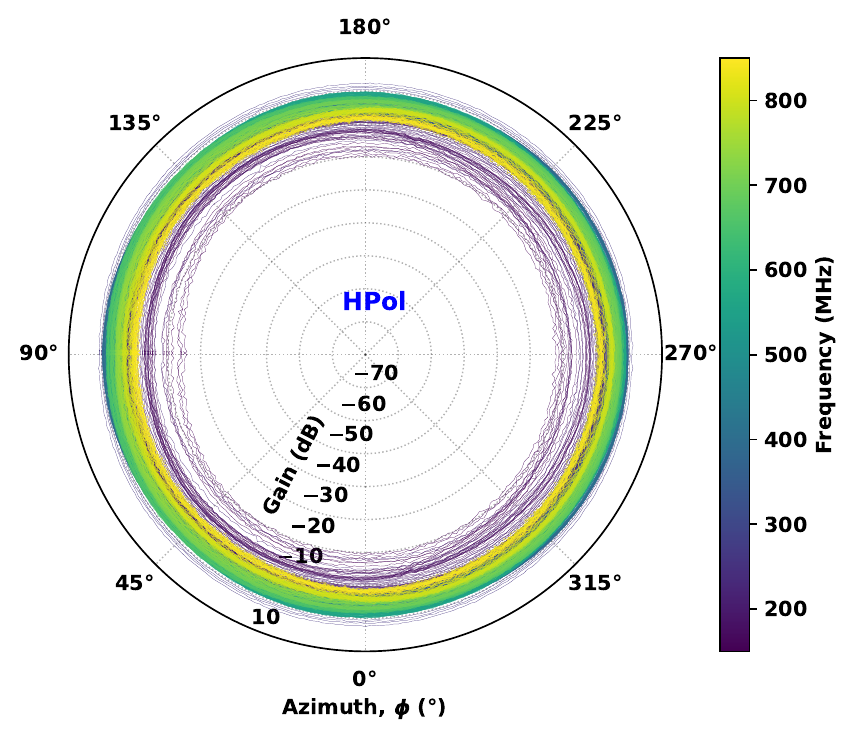}
        \caption{Realized gain patterns of ARA antennas in the zenith (left) and azimuthal (right) planes.}\label{fig5}
		\label{fig5}
\end{figure}

\subsection{Data Acquisition Center}\label{subsec2.3}
Entering the DAQ box, the signal is split and then enters the Triggering Daughter-board for ARA (TDA) and the Digitizing Daughter-board for ARA (DDA), which are mounted on the ARA Triggering and Readout Interface, or ATRI. Each station has 4 DDAs (DDA 0-3) and 4 TDAs (TDA 0-3), each connected to 4 antenna channels. An event is written to disk if the antenna signals exceed the  thresholds in at least 3 out of the 8 VPol or HPol antennas. Within each DDA, there is a digitizing chip called the Ice Ray Sampler second generation (IRS2) that records the data. Once the trigger condition is fulfilled, the IRS2 starts sampling and digitizing the data at a rate of 3.2 $\times 10^{9}$ samples/second with a low power consumption of 20 mW power per channel. The IRS2 chip has high power efficiency and fast sampling rate due to the Switched Capacitor Array architecture, comprising 128 capacitors arrays on two delay lines \cite{10} that help digitize 64 odd and 64 even samples in one odd-numbered and one even-numbered block. The analog data buffer can store 32k samples for 10 $\mu s$ which  corresponds to a total of $32\times 1024 = 32768$ samples at a rate of $32768/(10\times 10^{-6}) = 3.2$ GS/s. Each sample block saves up to 64 samples. A full readout, therefore, can save up to $32768/64 = 512$ blocks of data. The DAQ box also has GPS-synchronized Rubidium clock to trigger and control the CPs. Once data is digitized and saved, it is sent to the IceCube Lab hard disks and then to the North for storage on the University of Wisconsin - Madison's supercomputer cluster.

\section{Backgrounds}\label{sec3}
ARA, like most other UHEN and UHECR detectors, operates over a frequency range starting from $\mathcal{O}$(MHz) to $\mathcal{O}$(GHz) \cite{15}\cite{9}. There are numerous sources of backgrounds in this bandwidth, beginning with low frequency galactic noise \cite{16} to mid frequency radiosonde weather balloons  to unavoidable all-frequency thermal white-noise. We discuss a few such prominent backgrounds observed in ARA detectors.

\subsection{Thermal Noise}\label{subsec3.1}
Thermal noise is a low-power, constant background over the ARA bandwidth. It is temperature-dependent and is thus reduced in the cold \cite{18} South Pole environment. ARA Testbed surface antennas measured the thermal noise spectral density to be a bit higher than that at room temperature due to the added noise temperature contributed by its electronics \cite{9}. The in-ice ARA antenna temperature is about 247 K which is mostly contributed by ambient ice, bedrock, atmospheric, and galactic noise \cite{10}. ARA saves thermal noise at a 1 Hz rate to study the background and develop a data-driven noise model as shown in figure \ref{fig9}. 

\subsection{Continuous Waves}\label{subsec3.2}
\begin{figure}[h]
        \centering \includegraphics[scale = 0.18]{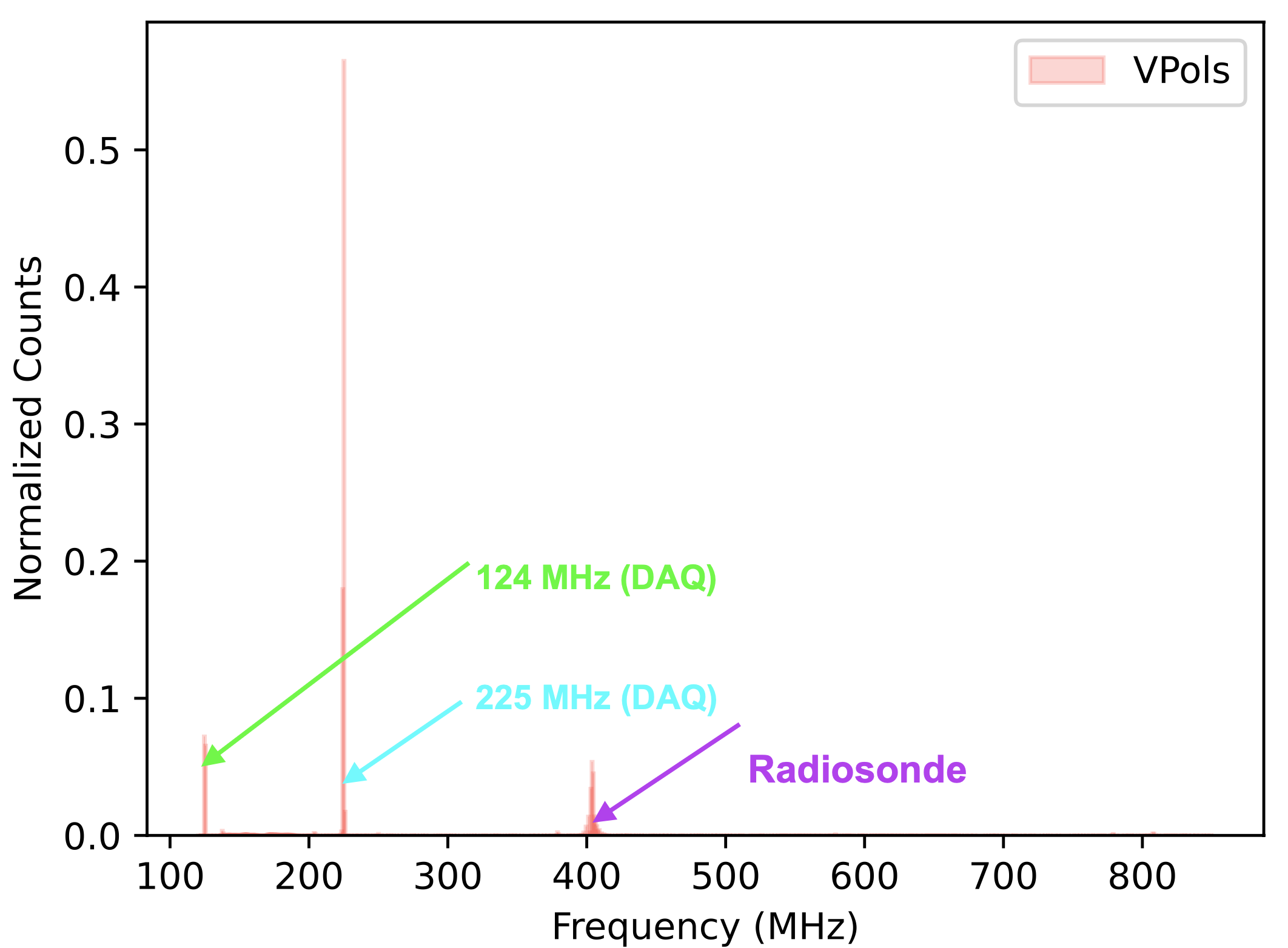}~\includegraphics[scale = 0.45]{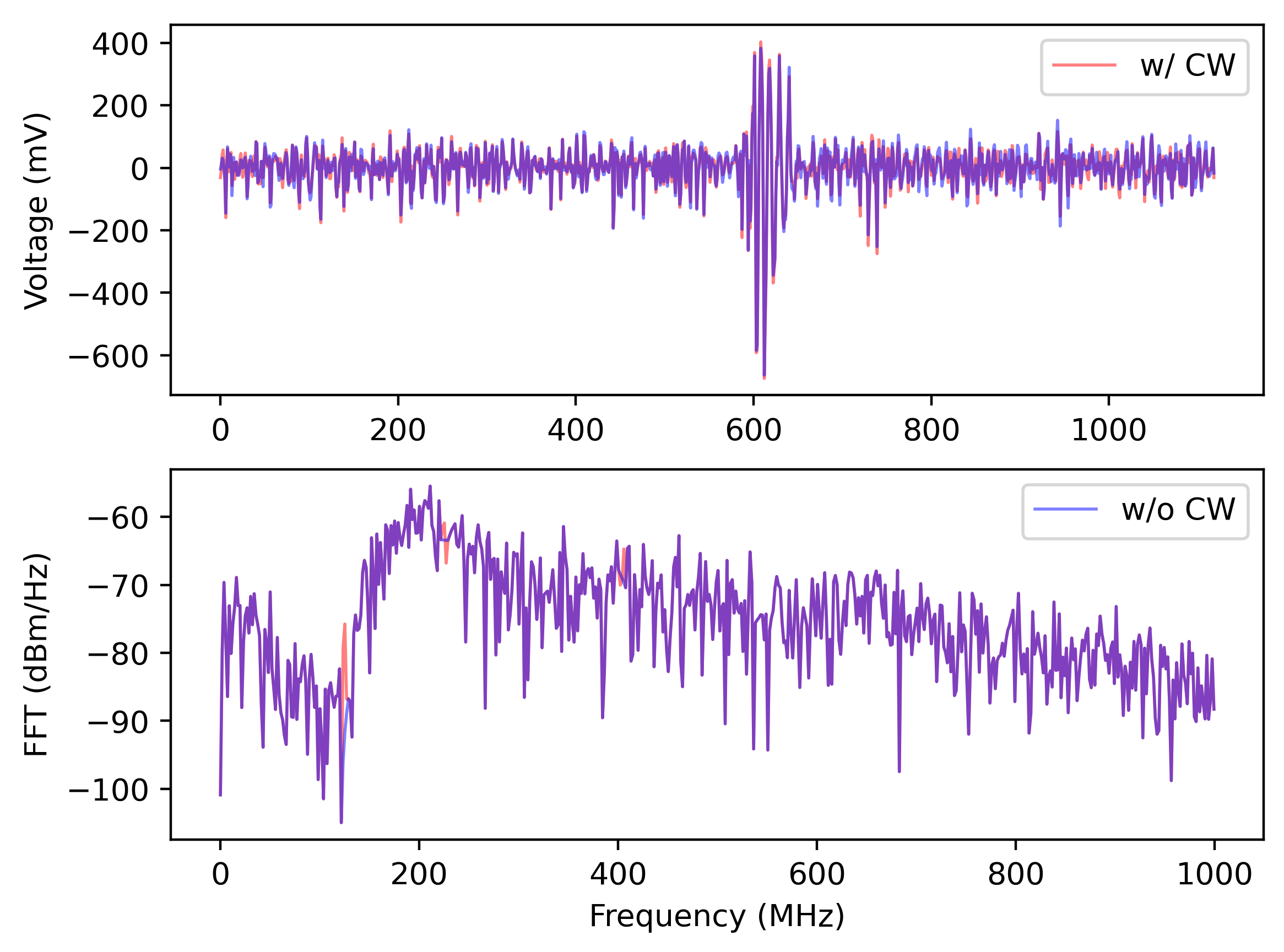}
		\caption{Some sources of CW backgrounds (left) in the VPol antennas and the effect of CW removal on waveforms (right top; x-axis is interpolated sample number) and their FFTs (right bottom).}
		\label{fig6}
\end{figure}

Continuous Waves (CW) are another prominent source of background in ARA mostly generated from the radiosonde NOAA weather balloons, at a frequency centered at 403 MHz \cite{17}. They can also be generated from the DAQ electronics, radio communications on the surface, and satellite links. A few CW sources are shown in figure \ref{fig6} (left). During data analysis, such frequencies are filtered out from the waveform. Figure \ref{fig6} (right) shows an A1 waveform and its power spectral density with (\textcolor{red!60!white}{w/}) and without (\textcolor{blue!60!white}{w/o}) CWs. 

\subsection{Transient Pulses}\label{subsec3.3}
In-ice Askaryan signals are transient and narrow-band \cite{6} and can be mimicked by other transient, pulsed signals. Such signals can be generated from in-air UHECR interactions, solar radio bursts \cite{9}\cite{21}, anthropogenic noise or even wind storms \cite{19} and represent a potential background in ARA. However, for a neutrino search, those backgrounds are filtered out by a surface correlation cut \cite{10}\cite{20}\cite{22}. 

\section{Past Analyses}\label{sec4}
ARA has been collecting data since early 2010 starting from the Testbed to five autonomous stations with a low threshold PA. Currently A3 and A5+PA are currently collecting data; other stations are awaiting a planned instrument upgrade in the upcoming Antarctic summer season. Although past data analyses searching for UHE cosmogenic neutrino ARA did not find a positive signal, world-best sensitivity allowed limits were set, and techniques defined for successor UHEN and UHECR experiments. 

\subsection{ARA Testbed}\label{subsec4.1}
\begin{figure}[h]
        \centering \includegraphics[scale = 0.033]{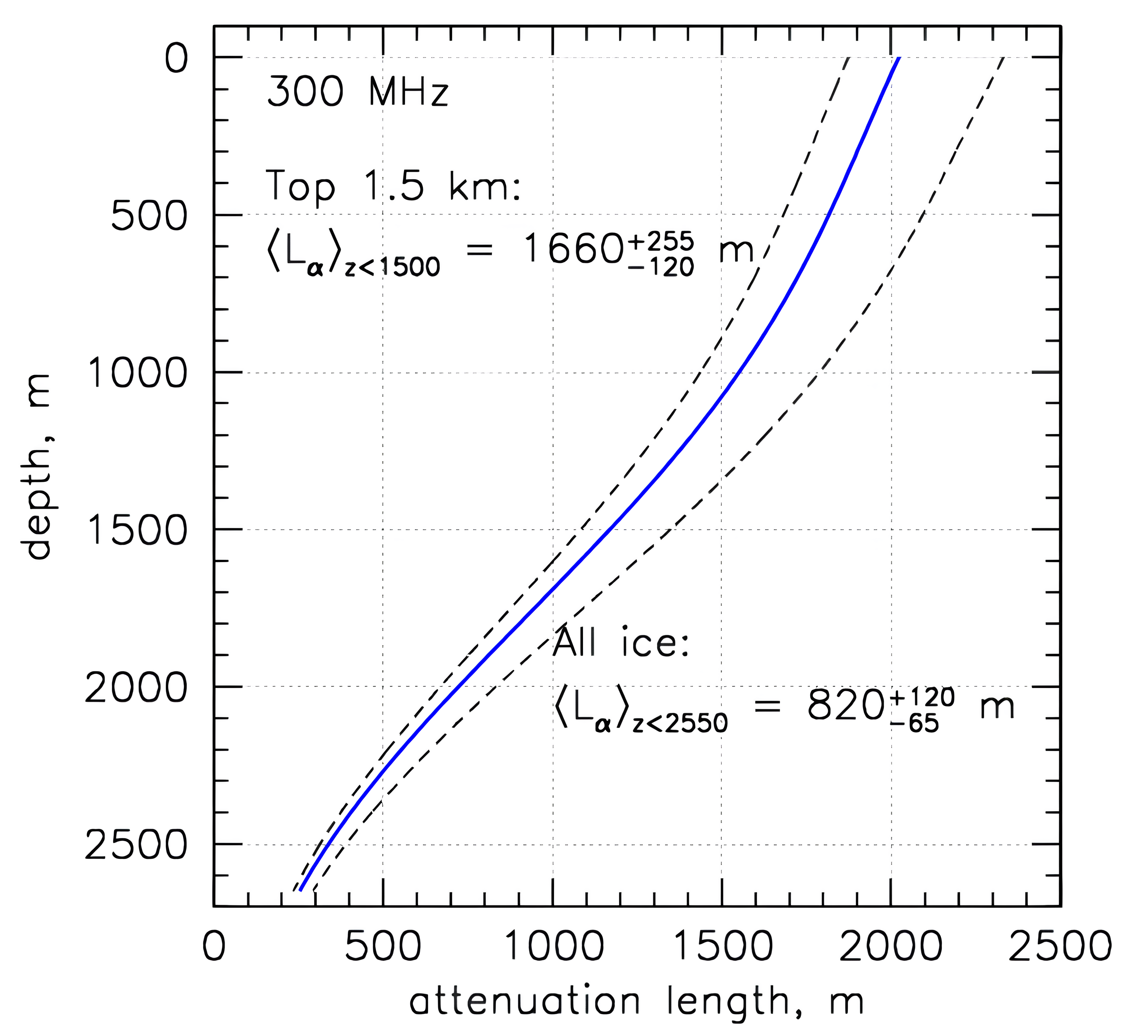}~\includegraphics[scale = 0.24]{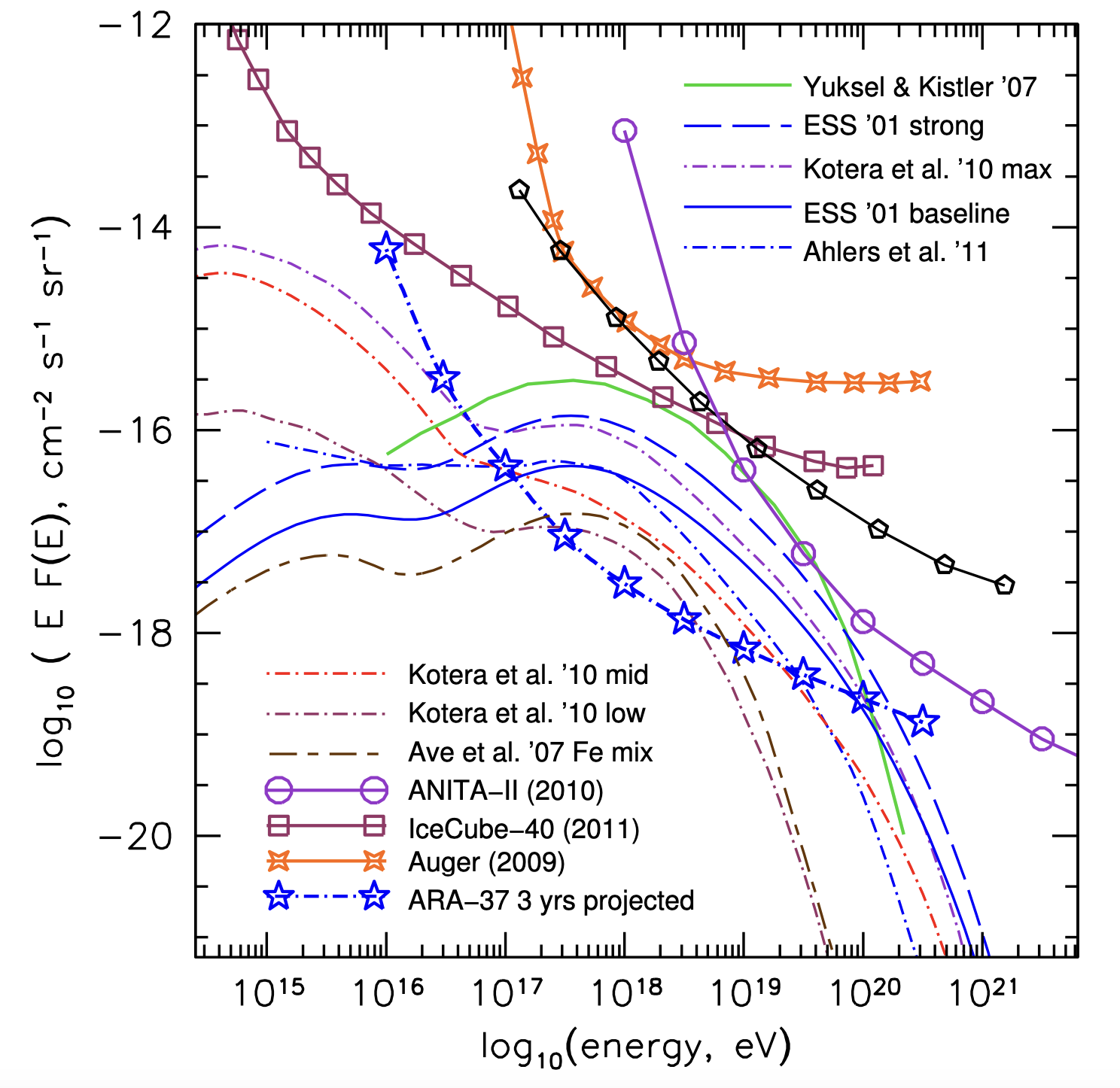}
		\caption{RF attenuation length (left) and projected sensitivity of 3 years of ARA-37 (right). Source \cite{9}.}
		\label{fig7}
\end{figure}

Following the successful testing and interesting outcomes of the RICE experiment, the ARA Testbed was deployed as a proof of concept for the then planned future ARA-37 \cite{9}. The testbed analyzed its full dataset and reported successful instrument testing by precisely measuring thermal floor, scanning the environment, and detecting Solar radio burst on February 13, 2011 during the previous solar maximum\cite{9}. In addition, it demonstrated reconstruction of the local CP with sub-degree precision and detected radio signals from distant deep pulser. It estimated the all-depth radio frequency average attenuation length to be $820^{+120}_{-65}$ m at 300 MHz. The Testbed analysis demonstrated real prospects for the detection of UHE cosmogenic neutrino in 3 years of ARA-37 data as shown in figure \ref{fig7}.

\subsection{A2+A3 Four Years}\label{subsec4.3}
\begin{figure}[h]
        \centering \includegraphics[scale = 0.24]{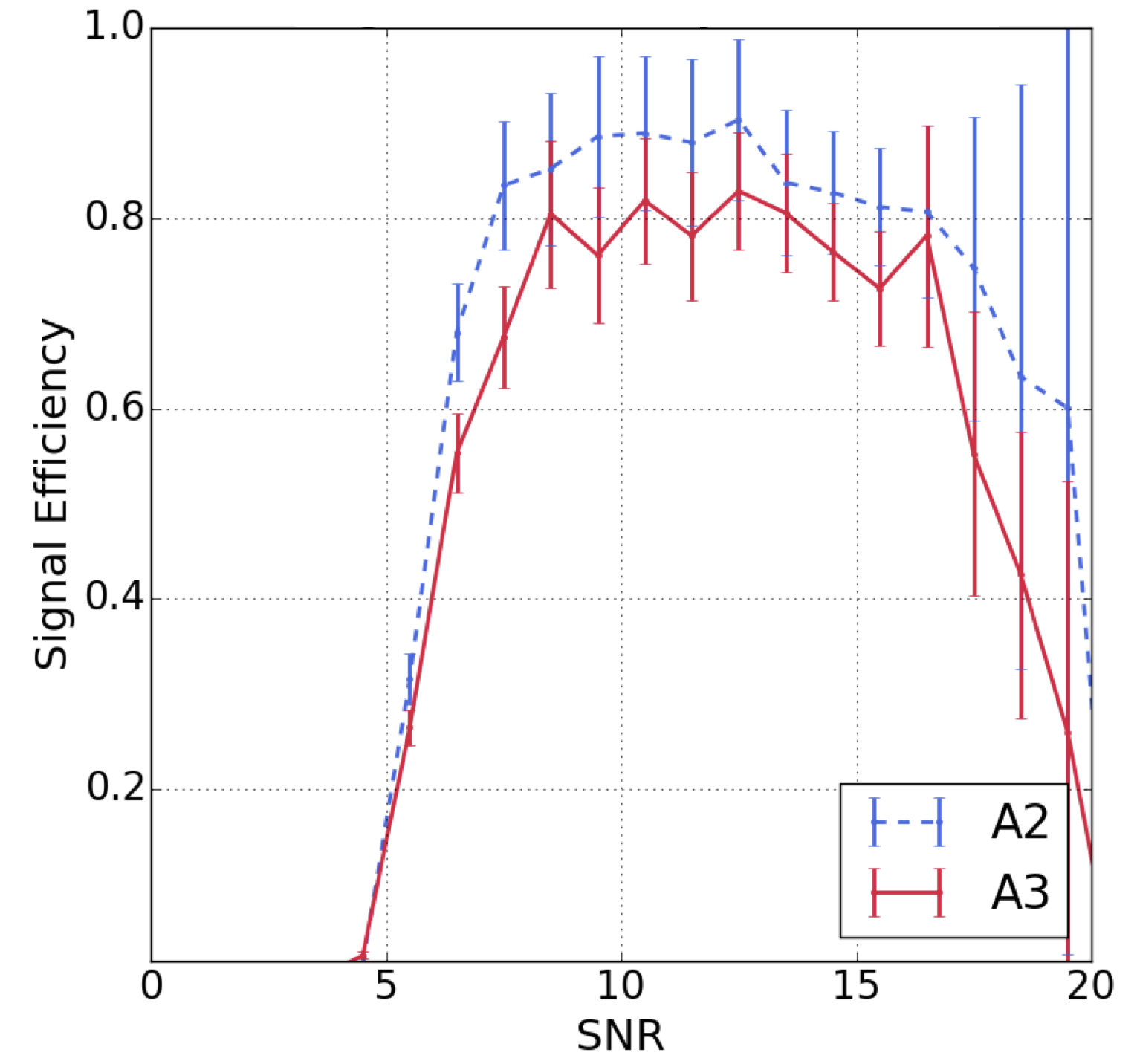}~\includegraphics[scale = 0.25]{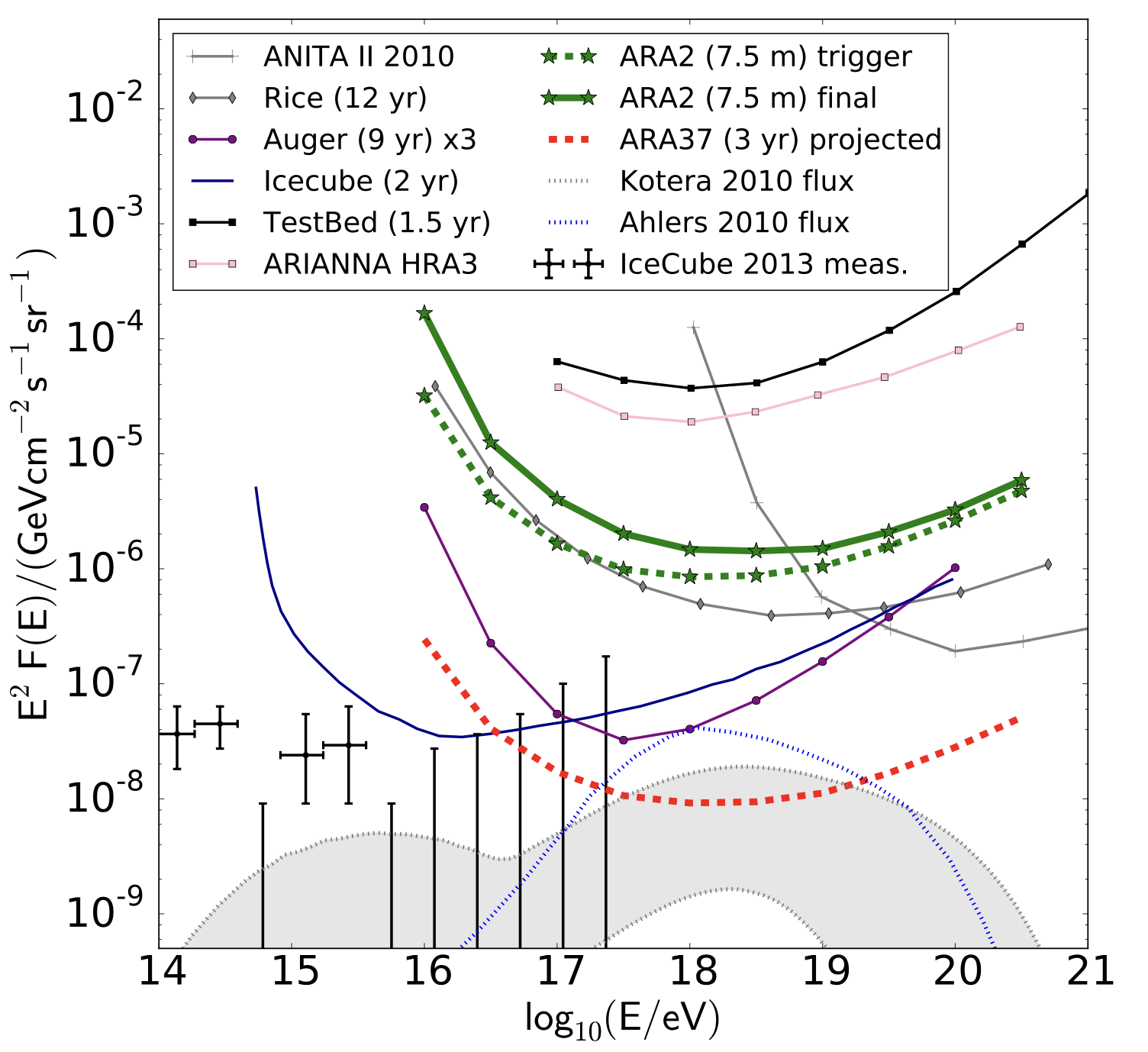}
		\caption{Signal efficiency vs SNR of A2 and A3 (left). Single-Event Sensitivity (SES) of 7.5 months of A2 data (right). Sources \cite{22}\cite{25}.}
		\label{fig8}
\end{figure}

By 2017, ARA stations 2 and 3 had accumulated a total of about six years (1141 days of A2 and 1021 days of A3) of data useful for a UHEN search. This analysis introduced several new techniques and identified a few important challenges to overcome in large scale radio neutrino experiments. Based on this search, they placed a trigger level 90\% CL upper limit, as shown in figure \ref{fig10}, on the EeV energy diffuse neutrino flux. They also predicted that, with a larger data set, ARA5's sensitivity will be world-leading above 100 PeV. An earlier search on A2+A3 was also conducted on 10 months of data which reported the single event sensitivity (SES) shown in figure \ref{fig8}, better than that of 1.5 years Testbed, indicating improved performance and larger detector volume. 

\subsection{PA 7 Months}\label{subsec4.2}
The design and deployment of an interferometric trigger phased array in the center of A5 \cite{26}\cite{20} ran at a  trigger rate to 11 Hz (with a correspondingly lower SNR threshold of 2.0), to be compared with 6 Hz (and a threshold of 3.7) in the conventional ARA stations. In 2022, a search was conducted using 7 months of PA data with the lowest signal threshold realized by any contemporary radio neutrino experiment. The PA paved the path for future large scale radio detectors \cite{15} to use phased arrays for lowering trigger thresholds and increasing the analysis-level detector effective volume. The sensitivity it produced with only 7 months of data, as shown in figure \ref{fig10}, was comparable with that of previous ARA searches and other experiments \cite{20}.

\section{ARA1-5 Analysis}\label{sec5}
ARA has been actively collecting data for about more than 10 years. Efforts are ongoing to incorporate the contributions from all 5 stations, including the phased array. This is the first array-wide analysis via global optimization across the entire array targeting discovery of UHEN. In addition to utilizing all accumulated data through 2023, ARA has made several changes in the analysis techniques and models. In this section, we discuss briefly on the status of the ARA1-5 analysis.
\subsection{Calibrations}\label{subsec5.1}
The IRS2 chip fabrication errors can induce a timing jitter of $\mathcal{O}$(100 ps) during sampling and digitization of data. Also, the recorded data format in ADC counts needs to be converted to voltage. In addition to timing jitter corrections and determining the ADC to voltage conversion factors, the locations of in-ice antennas need to be determined for vertex reconstruction of UHENs. ARA stations A1 \cite{12}, A2+A3 \cite{10}\cite{25} and A4+A5 \cite{13} have all been fully calibrated. Their pointing resolutions have been tested with local shallow CP and distant deep pulsers and have demonstrated sub-degree precision \cite{12}\cite{14}. 

\subsection{Data Driven Models}\label{subsec5.2}
\begin{figure}[h]
        \centering \includegraphics[scale = 0.26]{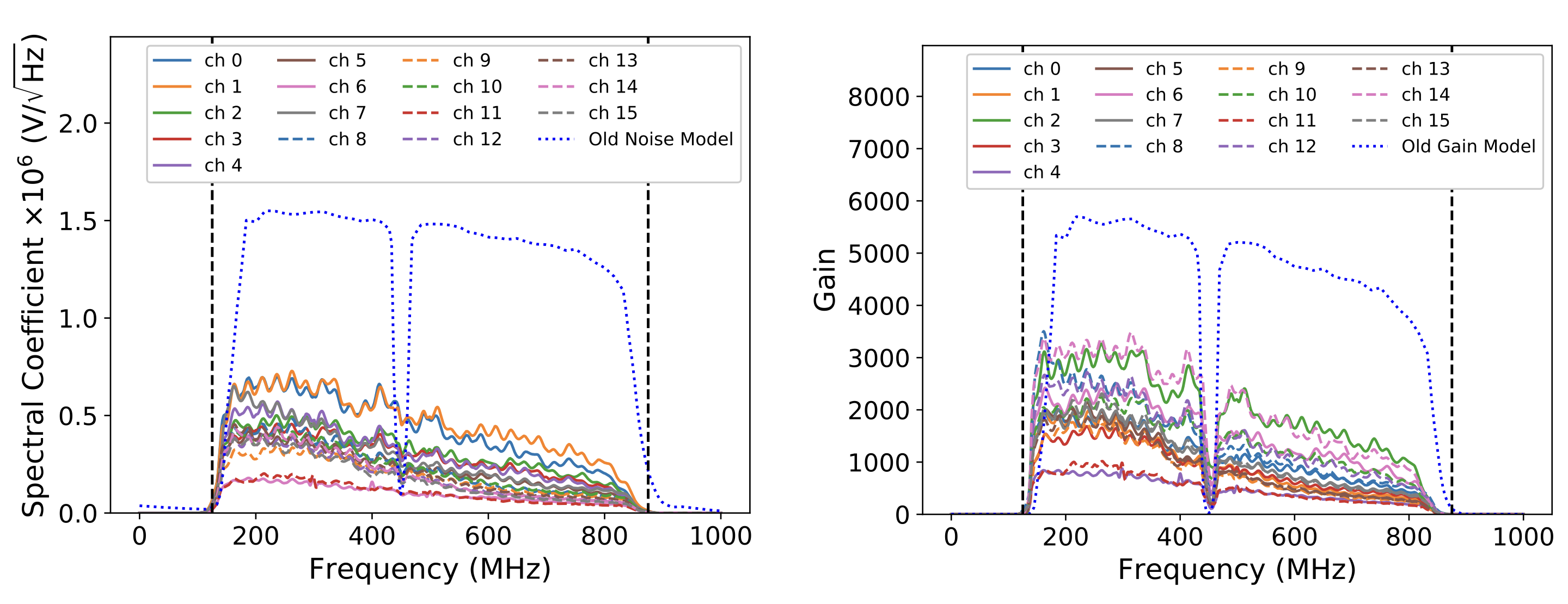}
		\caption{New data driven per channel noise (left) and gain (right) models compared to the previous models (blue dotted lines) within the ARA bandwidth (black dashed lines). Source \cite{11}.}
		\label{fig9}
\end{figure}

With the calibrations done, new data-driven-antenna-specific models have been developed. In previous analyses, SC gains and noise (blue dotted lines in figure \ref{fig9}) models have been used. For ARA1-5 analysis, a new set of noise and gain models have been developed for each antenna using the data from each station. The models specific to each station are subdivided in terms of the stations' detector configurations. Antennas, discussed in section \ref{subsec2.1}, are also modeled based on measurements at the University of Kansas anechoic chamber and then migrated, using the known ice refractive index. 

\subsection{Array-wide Neutrino Search}\label{subsec5.3}
\begin{figure}[h]
        \centering \includegraphics[scale = 0.225]{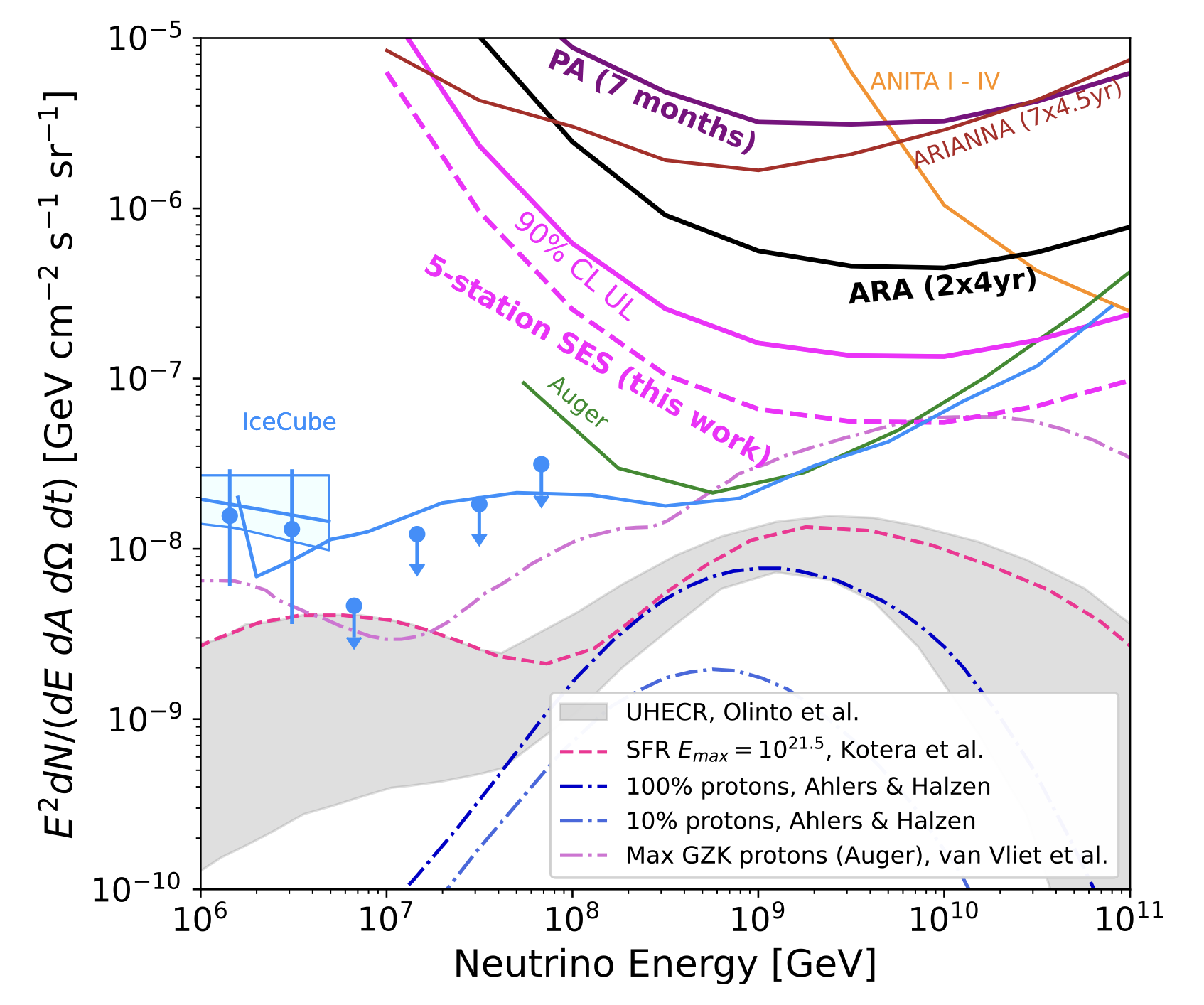}~\includegraphics[scale = 0.24]{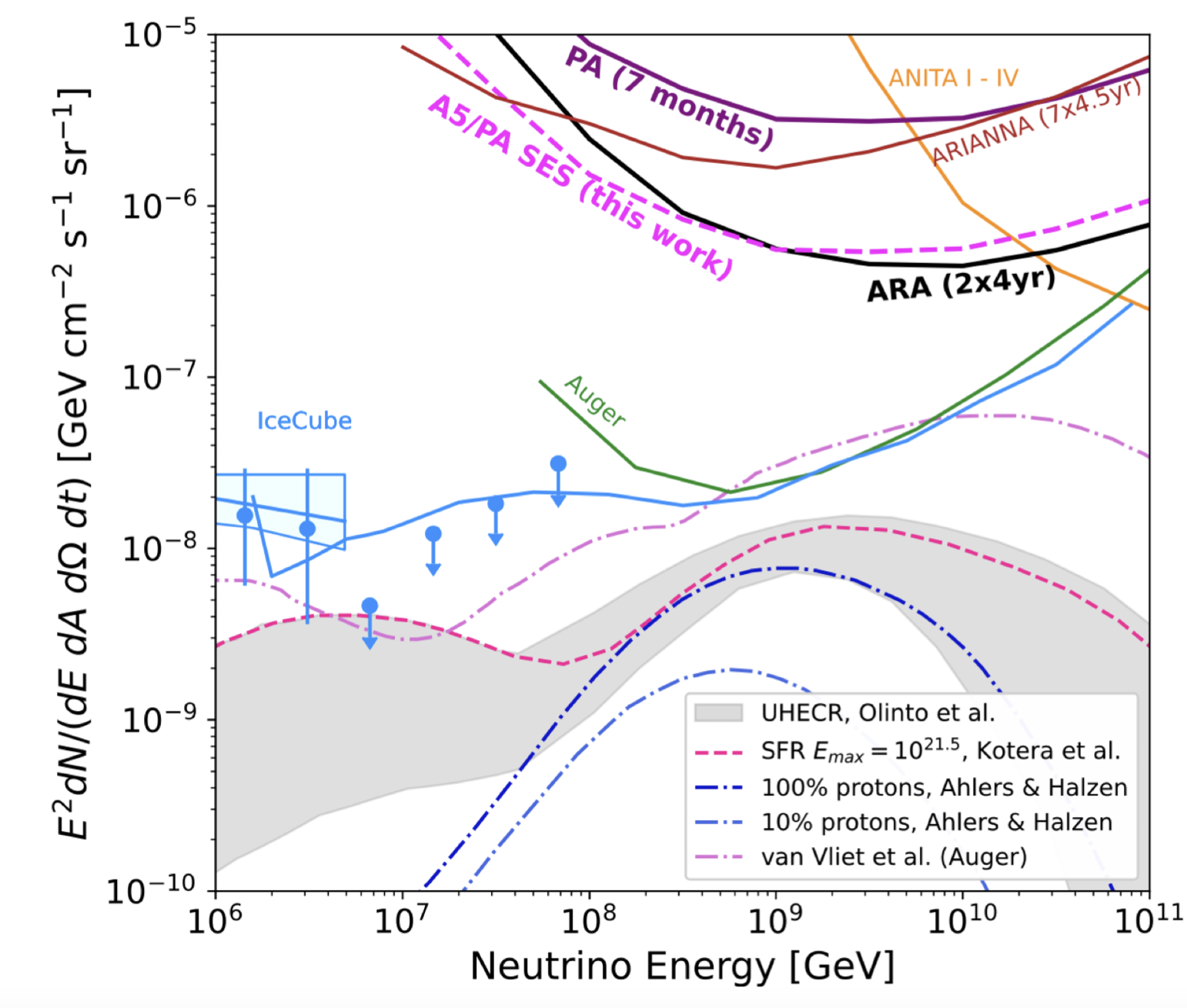}
		\caption{Projected Sensitivities of ARA1-5 (left) multi-array and A5+PA (right) hybrid searches. Also shown as the black line is the A2+A3 4 year search and, as the violet line, the PA search using 7 months of data. Sources \cite{11}\cite{23}.}
		\label{fig10}
\end{figure}

Through 2023, ARA has accumulated more than 27 station-years of data. The collaboration has invested adequate time to achieve good detector modeling. With that, and coordinating the efforts of multiple analyzers from multiple institutions, this will be the first of its kind search for UHEN. After station level analyses, a global optimization will be done on the linear discriminant values of all stations' analysis variables. Alongside the ARA1-5 analysis, a separate A5+PA hybrid analysis is also being conducted with improved models and analysis frameworks. The projected sensitivity shown in figure \ref{fig10} shows ARA's potential to detect the first cosmogenic neutrino. In the event of no signal detection, ARA will produce a world-leading upper limit on the diffuse neutrino flux below 1000 EeV neutrino energy \cite{11}. 

\section{Summary \& Outlook}\label{sec6}
ARA has been hunting for the highest energy neutrinos in the universe for more than a decade, starting from the Testbed, to ARA1-5, to the PA. In this article, we briefly discussed its science goals, detector designs, and potential backgrounds. We also highlighted a few past analyses searching for UHENs. Finally, we outlined an ongoing array-wide cosmogenic neutrino search. If UHEN are not detected, ARA will set the world's leading upper limit on the diffuse neutrino flux below $10^{12}$ GeV, in the radio-sensitive regime.\\\\
With the advancement of instruments and technology, new and more efficient radio neutrino detectors have evolved. ARA is also being planned to be upgraded with the new DAQ and trigger modules, ARA-Next \cite{24}. With future upgrades, improved techniques and models, and possible expansion, ARA holds significant potential to contribute towards multi-messenger astronomy.

\clearpage
\section*{Acknowledgements}

\noindent
The ARA Collaboration is grateful to support from the National Science Foundation through Award 2013134.
The ARA Collaboration
designed, constructed, and now operates the ARA detectors. We would like to thank IceCube, and specifically the winterovers for the support in operating the
detector. Data processing and calibration, Monte Carlo
simulations of the detector and of theoretical models
and data analyses were performed by a large number
of collaboration members, who also discussed and approved the scientific results presented here. We are
thankful to Antarctic Support Contractor staff, a Leidos unit 
for field support and enabling our work on the harshest continent. We thank the National Science Foundation (NSF) Office of Polar Programs and
Physics Division for funding support. We further thank
the Taiwan National Science Councils Vanguard Program NSC 92-2628-M-002-09 and the Belgian F.R.S.-
FNRS Grant 4.4508.01 and FWO. 
K. Hughes thanks the NSF for
support through the Graduate Research Fellowship Program Award DGE-1746045. A. Connolly thanks the NSF for
Award 1806923 and 2209588, and also acknowledges the Ohio Supercomputer Center. S. A. Wissel thanks the NSF for support through CAREER Award 2033500.
A. Vieregg thanks the Sloan Foundation and the Research Corporation for Science Advancement, the Research Computing Center and the Kavli Institute for Cosmological Physics at the University of Chicago for the resources they provided. R. Nichol thanks the Leverhulme
Trust for their support. K.D. de Vries is supported by
European Research Council under the European Unions
Horizon research and innovation program (grant agreement 763 No 805486). D. Besson, I. Kravchenko, and D. Seckel thank the NSF for support through the IceCube EPSCoR Initiative (Award ID 2019597). M.S. Muzio thanks the NSF for support through the MPS-Ascend Postdoctoral Fellowship under Award 2138121. A. Bishop thanks the Belgian American Education Foundation for their Graduate Fellowship support.

\bibliographystyle{unsrt}

\clearpage
{\small 
\section*{Full Author List: ARA Collaboration (July 22, 2024)}

\noindent
S.~Ali\textsuperscript{1},
P.~Allison\textsuperscript{2},
S.~Archambault\textsuperscript{3},
J.J.~Beatty\textsuperscript{2},
D.Z.~Besson\textsuperscript{1},
A.~Bishop\textsuperscript{4},
P.~Chen\textsuperscript{5},
Y.C.~Chen\textsuperscript{5},
Y.-C.~Chen\textsuperscript{5},
B.A.~Clark\textsuperscript{6},
A.~Connolly\textsuperscript{2},
K.~Couberly\textsuperscript{1},
L.~Cremonesi\textsuperscript{7},
A.~Cummings\textsuperscript{8,9,10},
P.~Dasgupta\textsuperscript{2},
R.~Debolt\textsuperscript{2},
S.~de~Kockere\textsuperscript{11},
K.D.~de~Vries\textsuperscript{11},
C.~Deaconu\textsuperscript{12},
M.~A.~DuVernois\textsuperscript{4},
J.~Flaherty\textsuperscript{2},
E.~Friedman\textsuperscript{6},
R.~Gaior\textsuperscript{3},
P.~Giri\textsuperscript{13},
J.~Hanson\textsuperscript{14},
N.~Harty\textsuperscript{15},
K.D.~Hoffman\textsuperscript{6},
M.-H.~Huang\textsuperscript{5,16},
K.~Hughes\textsuperscript{2},
A.~Ishihara\textsuperscript{3},
A.~Karle\textsuperscript{4},
J.L.~Kelley\textsuperscript{4},
K.-C.~Kim\textsuperscript{6},
M.-C.~Kim\textsuperscript{3},
I.~Kravchenko\textsuperscript{13},
R.~Krebs\textsuperscript{8,9},
C.Y.~Kuo\textsuperscript{5},
K.~Kurusu\textsuperscript{3},
U.A.~Latif\textsuperscript{11},
C.H.~Liu\textsuperscript{13},
T.C.~Liu\textsuperscript{5,17},
W.~Luszczak\textsuperscript{2},
K.~Mase\textsuperscript{3},
M.S.~Muzio\textsuperscript{8,9,10},
J.~Nam\textsuperscript{5},
R.J.~Nichol\textsuperscript{7},
A.~Novikov\textsuperscript{15},
A.~Nozdrina\textsuperscript{1},
E.~Oberla\textsuperscript{12},
Y.~Pan\textsuperscript{15},
C.~Pfendner\textsuperscript{18},
N.~Punsuebsay\textsuperscript{15},
J.~Roth\textsuperscript{15},
A.~Salcedo-Gomez\textsuperscript{2},
D.~Seckel\textsuperscript{15},
M.F.H.~Seikh\textsuperscript{1,*},
Y.-S.~Shiao\textsuperscript{5,19},
S.C.~Su\textsuperscript{5},
S.~Toscano\textsuperscript{20},
J.~Torres\textsuperscript{2},
J.~Touart\textsuperscript{6},
N.~van~Eijndhoven\textsuperscript{11},
G.S.~Varner\textsuperscript{21,$\dagger$},
A.~Vieregg\textsuperscript{12},
M.-Z.~Wang\textsuperscript{5},
S.-H.~Wang\textsuperscript{5},
S.A.~Wissel\textsuperscript{8,9,10},
C.~Xie\textsuperscript{7},
S.~Yoshida\textsuperscript{3},
R.~Young\textsuperscript{1}
\\
\\
\textsuperscript{$*$} Corresponding author\\
\textsuperscript{$\dagger$} Deceased\\
\\
\textsuperscript{1} Dept. of Physics and Astronomy, University of Kansas, Lawrence, KS 66045\\
\textsuperscript{2} Dept. of Physics, Center for Cosmology and AstroParticle Physics, The Ohio State University, Columbus, OH 43210\\
\textsuperscript{3} Dept. of Physics, Chiba University, Chiba, Japan\\
\textsuperscript{4} Dept. of Physics, University of Wisconsin-Madison, Madison,  WI 53706\\
\textsuperscript{5} Dept. of Physics, Grad. Inst. of Astrophys., Leung Center for Cosmology and Particle Astrophysics, National Taiwan University, Taipei, Taiwan\\
\textsuperscript{6} Dept. of Physics, University of Maryland, College Park, MD 20742\\
\textsuperscript{7} Dept. of Physics and Astronomy, University College London, London, United Kingdom\\
\textsuperscript{8} Center for Multi-Messenger Astrophysics, Institute for Gravitation and the Cosmos, Pennsylvania State University, University Park, PA 16802\\
\textsuperscript{9} Dept. of Physics, Pennsylvania State University, University Park, PA 16802\\
\textsuperscript{10} Dept. of Astronomy and Astrophysics, Pennsylvania State University, University Park, PA 16802\\
\textsuperscript{11} Vrije Universiteit Brussel, Brussels, Belgium\\
\textsuperscript{12} Dept. of Physics, Enrico Fermi Institue, Kavli Institute for Cosmological Physics, University of Chicago, Chicago, IL 60637\\
\textsuperscript{13} Dept. of Physics and Astronomy, University of Nebraska, Lincoln, Nebraska 68588\\
\textsuperscript{14} Dept. Physics and Astronomy, Whittier College, Whittier, CA 90602\\
\textsuperscript{15} Dept. of Physics, University of Delaware, Newark, DE 19716\\
\textsuperscript{16} Dept. of Energy Engineering, National United University, Miaoli, Taiwan\\
\textsuperscript{17} Dept. of Applied Physics, National Pingtung University, Pingtung City, Pingtung County 900393, Taiwan\\
\textsuperscript{18} Dept. of Physics and Astronomy, Denison University, Granville, Ohio 43023\\
\textsuperscript{19} National Nano Device Laboratories, Hsinchu 300, Taiwan\\
\textsuperscript{20} Universite Libre de Bruxelles, Science Faculty CP230, B-1050 Brussels, Belgium\\
\textsuperscript{21} Dept. of Physics and Astronomy, University of Hawaii, Manoa, HI 96822\\
}

\end{document}